%% file: references.tex
\documentclass[manuscript,screen]{acmart}

\AtBeginDocument{%
  \providecommand\BibTeX{{%
    \normalfont B\kern-0.5em{\scshape i\kern-0.25em b}\kern-0.8em\TeX}}}

\setcopyright{acmcopyright}
\copyrightyear{2023}
\acmYear{2023}
\acmDOI{}

\acmConference[CHI'23]{Proceedings of the Generative AI and HCI Workshop at the ACM Conference on Human Factors in Computing Systems}{April 23--28}{Hamburg, Germany}
\acmBooktitle{Proceedings of the Generative AI and HCI Workshop at the ACM Conference on Human Factors in Computing Systems, April 23--28, 2023, Hamburg, Germany} 

\usepackage{color, colortbl}

\begin{document}

\title{Power-up! What Can Generative Models Do for Human Computation Workflows?}


\author{Garrett Allen}
\email{G.M.Allen@tudelft.nl}
\orcid{0000-0003-4449-1510}
\author{Gaole He}
\email{G.He@tudelft.nl}
\orcid{0000-0002-8152-4791}
\author{Ujwal Gadiraju}
\orcid{0000-0002-6189-6539}
\email{U.K.Gadiraju@tudelft.nl}
\affiliation{%
  \institution{Delft University of Technology}
  \city{Delft}
  \state{Zuid-Holland}
  \country{Netherlands}
}

\renewcommand{\shortauthors}{{Allen, et al.}}

\begin{abstract}
  We are amidst an explosion of artificial intelligence research, particularly around large language models (LLMs). These models have a range of applications across domains like medicine, finance, commonsense knowledge graphs, and crowdsourcing. Investigation into LLMs as part of crowdsourcing workflows remains an under-explored space. The crowdsourcing research community has produced a body of work investigating workflows and methods for managing complex tasks using hybrid human-AI methods. Within crowdsourcing, the role of LLMs can be envisioned as akin to a cog in a larger wheel of workflows. From an empirical standpoint, little is currently understood about how LLMs can improve the effectiveness of crowdsourcing workflows and how such workflows can be evaluated. In this work, we present a vision for exploring this gap from the perspectives of various stakeholders involved in the crowdsourcing paradigm --- the task requesters, crowd workers, platforms, and end-users. We identify junctures in typical crowdsourcing workflows at which the introduction of LLMs can play a beneficial role and propose means to augment existing design patterns for crowd work.
\end{abstract}




\maketitle

\input{sections/sec-introduction}
\input{sections/sec-vision}

\bibliographystyle{ACM-Reference-Format}
\bibliography{references}

\appendix

\end{document}

%% file: sections/sec-introduction.tex
\section{Introduction and Background}

Artificial intelligence (AI) research is being reinvigorated with current advances in large language models (LLMs). Since their inception, LLMs have increased in size, effectiveness, and applications. For instance, BERT~\cite{devlin2018bert}, initially trained for masked language prediction, has been applied to other domains such as neural ranking~\cite{han2020learning,nogueira2019passage} and document classification~\cite{adhikari2019docbert,kong2022hierarchical}. OpenAI's\footnote{\url{https://www.openai.com/}} GPT family of models have been used in language tasks including goal-oriented dialogue~\cite{ham2020end}, patent claim generation~\cite{lee2020patent}, and story generation~\cite{lucy2021gender}. The most recent GPT variant, ChatGPT~\cite{openai2022chatgpt}, has seen an explosive growth in popularity, indicating the potential for a promising future where LLMs are deployed as work assistants. Due to such powerful generative capability, more researchers have started exploring generative LLMs in work assistant roles. For example, powerful generative LLMs have shown human-comparable writing skills in story generation~\cite{yuan2022wordcraft} and scientific writing~\cite{gero2022sparks}. LLMs have also exhibited promising assistive capability in complex tasks like coding~\cite{friedman_2021}, drug discovery~\cite{liu2021ai}, and question generation for education needs~\cite{wang2022towards}.

The common thread running through all variations in LLMs is the need of high quality data for training and evaluation. Crowdsourcing has been widely adopted in machine learning practice to obtain high-quality annotations by relying on human intelligence~\cite{vaughan2017making,gadiraju2020can}. Crowdsourcing is a paradigm in which researchers or other stakeholders request the participation of a distributed crowd of individuals, who can contribute with their knowledge, expertise, and experience~\cite{estelles2012towards}. Such individuals, called \textit{crowd workers}, are asked to complete a variety of tasks in return for monetary or other forms of compensation. Tasks are often decomposed into smaller atomic units and can vary in their purpose, including labelling images, editing text, or finding information on specific topics~\cite{gadiraju2014taxonomy}. Tasks can be standalone, or organized as a series of smaller sub-tasks, depending on their overall complexity and the design choices made by requesters. More complex problems, such as software engineering or system design problems, require task workflows.

\textit{Crowdsourcing workflows} are distinct patterns that manage how large-scale problems are decomposed into smaller tasks to be completed by workers. The crowd-powered word processor Soylent applies the \textit{Find-Fix-Verify} workflow to produce high-quality text by separating tasks into generating and reviewing text~\cite{bernstein2010soylent}. The \textit{Iterate-and-Vote} workflow has been deployed in creating image descriptions, where workers are asked to write descriptions of images to assist those who are blind~\cite{little2010exploring}. Subsequent voting tasks are used to decide on the optimal description.~\citet{chen2015crowdmr} introduce CrowdMR, which combines the \textit{Map-Reduce} workflow with crowdsourcing to facilitate the solving of problems that require both human and machine intelligence, i.e., ``AI-Hard" problems~\cite{yampolskiy2012ai}. With CrowdForge,~\citet{kittur2011crowdforge} provide a framework for crowdsourcing to support complex and interdependent tasks. The authors follow up with the tool CrowdWeaver~\cite{kittur2012crowdweaver} for managing complex workflows, supporting such needs as data sharing between tasks and providing monitoring tools and real-time task adjustment capability. Taking a more holistic look at workflows,~\citet{retelny2017no} investigate the relationship between the need for adaptation and complex workflows within crowdsourcing, finding that the current state of crowdsourcing processes are inadequate for providing the necessary adaptation that complex workflows require. 

Within crowdsourcing, the role of LLMs can be envisioned as akin to a cog in a larger workflow. Typically, LLMs are used for supporting individual writing or classification tasks within a workflow, as previous examples expressed. Researchers are also exploring the application of LLMs in assisting crowd workers.~\citet{liu2022wanli} combine the generative power of GPT-3 and the evaluative power of humans to create a new natural language inference dataset that produces more effective models when used as a training set. In a similar vein,~\citet{bartolo2021models} introduce a ``Generative Annotation Assistant" to help in the production of dynamic adversarial data collection, significantly improving the rate of collection. These works measure the effectiveness of the models and the individual tasks, yet there remains an open gap regarding the understanding of how LLMs improve the effectiveness of crowdsourcing workflows and how such workflows can be evaluated.

In this work, we present a vision for exploring the gap from the stakeholders' perspectives, e.g., task requesters, crowd workers, and end-users. In so doing, we highlight the junctures of crowdsourcing workflows at which introducing LLMs can be beneficial. We also propose means to augment existing design patterns for crowd work.

%% file: sections/sec-vision.tex
\section{Incorporating Large Language Models in Crowdsourcing Workflows}

As LLMs are pre-trained on large text corpora, they show great capability in understanding context-specific semantics. When further fine-tuned for specific uses with additional, smaller datasets, highly effective and domain-targeted models can be produced. Additionally, some LLMs (e.g., BART~\cite{lewis2019bart}, GPT-3~\cite{brown2020language}) are also good at generating responses to input queries, which can be fluent, human-like, and even professional. As it stands, LLMs have been effectively deployed within multiple domains such as medicine~\cite{alsentzer2019publicly}, finance~\cite{yang2020finbert}, and others requiring commonsense reasoning~\cite{bosselut2019comet}. As such, LLMs are an opportune and potentially very useful tool to use within crowdsourcing where domain knowledge may not always be available.

While LLMs are effective in many ways, they are far from being perfect and come with drawbacks. Due to their black box neural backbone, LLMs suffer from a lack of transparency, which leads to difficulty in explaining how they achieve the performances they do~\cite{zini2022explainability}. Such opacity also makes it difficult to track the factual error of LLMs, which inhibits the potential for improving the models~\cite{dong2022calibrating}. Further, language models are known to capture several different forms of biases~\cite{abid2021persistent,nadeem2020stereoset,vig2020investigating}. Most existing LLMs tend to perform poorly on tasks that require commonsense knowledge~\cite{rajani2019explain}, which is a common practice for children. Last but not least, current language models achieve poor compositional generalization~\cite{hosseini2022compositional}, which is required for solving complex tasks. Noticeably, LLMs fall short in aspects that humans are good at, e.g., commonsense reasoning~\cite{li2021language} and complex task planning~\cite{ahn2022can,wei2022chain}. Putting LLMs into practice requires either addressing or working within these limitations.
    
\subsection{The Lens of Complex Crowdsourcing Workflows}

LLMs can easily fit into existing crowdsourcing workflows. Take the Find-Fix-Verify workflow as an example. This workflow is well-suited for writing tasks, whether it be editing, revisions, or new content. Each step is an opportunity to include LLMs for improvements in the process. Let us take the example of revising a news story. During the ``Find" stage, a workers would be tasked with reading the story and finding any errors, e.g., grammar, spelling, or false statements. Once these errors are identified, a new crowd of workers is recruited for the next stage: to ``Fix" the errors. We are now left with an updated draft of the news story that has fewer errors than the initial draft. Which brings us to the final stage of the workflow, ``Verify", where yet another group of workers validate the work of the prior groups. In this particular example, it is fairly clear where an LLM can be swapped for the workers at each stage. A retrieval or error classification LLM can be deployed for finding the errors, a generative LLM can be used to produce repaired text, and yet another classification LLM can finish it all off as the verifier. However, not all tasks take this form, or follow this particular workflow. Adapting other workflows, i.e., Iterate-and-Vote or MapReduce, can be done in a similar manner. Even so, adaptations such as these prompt the question: Once introduced, what are the effects of LLMs within crowdsourcing workflows for each stakeholder of the crowdsourcing process?

On the surface, this appears like straightforward question. Crowdsourcing has many different stakeholders involved: the \textit{requesters}, the \textit{workers}, and the \textit{end-users}. The impact of including an LLM into workflows has the potential to affect each stakeholder in different ways. From the perspective of the requester, the monetary cost of completing tasks will be reduced as potentially fewer workers will need to be recruited. The tasks may take less time to complete which will result in further monetary savings. A reduction in time to gather data, complete tasks, and/or a reduced need for workers may have a negative impact on the income flow for workers, however. With available tasks taking less time and there being fewer tasks, it creates the potential for crowd workers to earn less. This can be offset by adjusting incentive structures on platforms. On the other hand, the reduction in costs for requesters could lead to more tasks being posted, leading to more high-quality labels. In turn, LLMs benefit from the better labels and improve in performance as well, creating a positive cycle that benefits both crowd workers and requesters. Further work is required to gain a better understanding of the financial opportunities and risks surrounding LLMs as part of crowdsourcing workflows.

Of course, there are trade-offs that come alongside any benefits. The trade-off for the requesters is a learning curve around the LLMs. Time will need to be dedicated to strategize and familiarize with the integration of LLMs in workflows. A trade-off that crowdsourcing platforms will share, accompanied by the additional cost of the development to add the LLMs to their products. An LLM must be trained before it can be appropriately used within a crowdsourcing workflow. This training, or fine-tuning, creates an overhead for either the crowdsourcing platform or the requester. While the overhead is initially a burden for most stakeholders, there will be an efficiency gain in the long term. 

\subsection{Risk and Opportunity}
Further consideration is needed regarding the transparency of LLMs versus humans. When crowd workers complete tasks, such as annotation or other decision-oriented varieties, requesters have the capability of performing a follow-up with the workers to elicit reasoning for the outcomes provided. This is not a simple job for LLMs. While there exist methods for model explainability \cite{ribeiro2016should,ribeiro2018anchors,wu2021polyjuice}, none have demonstrated a level of effectiveness on par with what a requester would achieve with a human-human conversation. This same lack of transparency also has the potential of confounding workflows at the worker level. For example, take a scenario where an LLM is tasked with making a prediction, and a human worker to validate the prediction of the model, and the model provides a prediction that is not in line with what the worker expects to see. In such a scenario, the worker may want to interrogate the model to gain insight into why the prediction was made. However, there is currently no such clear way for the worker to request such an explanation from the LLM.

Also worth considering is the concept of accountability. Whenever a machine is introduced into a system, be it a factory, an airplane, or a crowdsourcing workflow, the question of accountability requires definition. Adding LLMs into crowdsourcing workflows raises the question of who or what is accountable if things do not go according to plan? Is the model, the requester, the platform, or the crowd workers to be held responsible for mishaps? There are many questions around the benefits, viability, risks, and harms involved with introducing LLMs into crowdsourcing workflows. These questions provide rich research opportunities for the generative AI and human computation research communities.

The realm of creative crowdsourcing tasks is another place of opportunity for LLMs. Generative models can help by providing suggestions or starting points to spark brainstorming or idea generation sessions. Alternatively, classification LLMs can be used to consolidate the ideas produced. For tasks that are more engineering or design focused, LLMs may be able to serve as ``rubber duck" sounding boards. LLMs may also provide performance boosts in areas such as content creation, music composition, or protein discovery. The possibilities of how LLMs can be included in crowdsourcing are vast, yet the viability of these use cases warrants further investigation.